# OPTIMIZATION OF PIEZOELECTRIC ELECTRICAL GENERATORS POWERED BY RANDOM VIBRATIONS

*Elie LEFEUVRE, Adrien BADEL, Claude RICHARD, Lionel PETIT, Daniel GUYOMAR*

Laboratoire de Génie Electrique et Ferroélectricité, INSA de Lyon,
bat. G. Ferrié, 8 rue de la physique, 69621 Villeurbanne Cedex, France

## ABSTRACT

This paper compares the performances of a vibration-powered electrical generators using PZT piezoelectric ceramic associated to two different power conditioning circuits. A new approach of the piezoelectric power conversion based on a nonlinear voltage processing is presented and implemented with a particular power conditioning circuit topology. Theoretical predictions and experimental results show that the nonlinear processing technique may increase the power harvested by a factor up to 4 compared to the Standard optimization technique. Properties of this new technique are analyzed in particular in the case of broadband, random vibrations, and compared to those of the Standard interface.

## 1. INTRODUCTION

The proliferation of transducers and sensors integrated in many systems raises the problem of of the installation of wires for power supplies and data transmission. In many applications, the limited lifespan of batteries may induce a costly maintenance. Moreover, batteries dying without warning poses serious problems in safety monitoring applications. This explains the growing interest in miniature electrical generators enabling to power transducers and their associated wireless data transmission systems. Sustainable power generation may be achieved in converting ambient energy into electrical energy. Some possible ambient energy sources are, for instance, thermal energy, light energy or mechanical energy. The focus here is on the transformation of mechanical vibrations, existing almost everywhere. In this field, electromagnetic and electrostatic generators have been developed. However, piezoelectric generators are of major interest due to their solid state nature facilitating their integration [1]. That's why different approaches of energy harvesting using piezoelectric materials have been developed over the past few years.

Optimization of Piezoelectric Electrical Generators (PEG) is an essential stage to consider their miniaturization while preserving a significant output power level. Previous works proposed a quasi-linear approach to maximize the PEG output power [2,3]. Although the power electronic interface used for optimization induces a distortion on the piezoelectric voltage, the technique is similar to an impedance adaptation between the load and the piezoelectric element [4]. More recently, techniques allowing to significantly enhancing the electromechanical energy conversions capability of piezoelectric materials were developed [5,6,7]. These techniques are based on nonlinear processing of the piezoelectric voltage synchronized with the mechanical vibrations. Compared to the quasi-linear impedance adaptation, theoretical and experimental results showed that these new techniques may increase the PEG power by a factor above 10 [8].

In addition to increasing the PEG power, some of these new nonlinear techniques intrinsically solve the problem of the matching load impedance [7,8]. This property is particularly interesting to optimize the energy harvested from broadband vibrations. Indeed, the matching load impedance required by the other optimisation techniques remains difficult to tune effectively in the case of broadband or random vibrations. This case is the most difficult to solve, but it is also the most frequent operating condition for PEGs: ambient vibrations are usually neither of constant amplitude nor on a single frequency. This paper analyses the broadband effectiveness of a PEG using a nonlinear self-tuned optimization technique with mechanical vibrations in the 50 Hz to 1 kHz region.

The theoretical results are experimentally verified for a PEG prototype composed of a piezoelectric ceramic bonded on a cantilever steel beam, subjected to harmonic and random, broadband forces.

## 2. ARCHITECTURES

A PEG is usually composed of a mechanical part, whose role is to collect ambient mechanical vibration and subject the active material to stress and strain variations. Through its electromechanical coupling property, the active material (for instance a PZT ceramic) converts the mechanical energy resulting from stress and strain variations into electrical energy. In open circuit configuration, an alternating voltage thus develops across the terminal electrodes of the active material.

Most of the electric loads are powered with a DC voltage, that's why an electronic power conversion interface is usually placed between the terminal electrodes of the active material and the power supply inputs of the electric load (Figure 1). This power conversion interface can be as simple as a diode rectifier associated to a voltage smoothing capacitor. However, from the electrical side, a piezoelectric element can be basically modeled by a current source in parallel with a capacitor (or a voltage source in series with the same capacitor) and the search for a maximization of the power leads naturally to a linear impedance adaptation approach [2,3,4,8]. From its power supply inputs, the electric load may be equivalent to a resistor, or a voltage source if a battery is used as energy storage. So, the maximum of power transmitted by the active material can be tracked by adapting the load current respectively with the variations of ambient vibrations. The power conversion interface may thus play a key role in PEGs power optimization.

## 2.1. Mechanical part

Many mechanical structures have been envisioned to transmit environmental vibrations to active material. These structures can be classified into two main categories. The first category includes the structures sensitive to environmental mechanical acceleration. In this case, the structures are mechanical resonators, having only one or several resonance frequencies. Effective transmission of surrounding vibrations to piezoelectric material thus depends on the matching between the resonance frequencies and the environmental frequencies. And only a few specific frequencies must be considered for power optimization. Also, mechanical behavior of the resonator may be significantly influenced by the energy conversion achieved by the piezoelectric material.

In the second category, PEGs mechanical structures directly use strain variations of the vibrating host structure and adapt it to active material. Instead of being limited to specific resonance frequencies, the frequency spectrum of mechanical strain/stress transmitted to the active material is a function of the strain variations existing in the host structure. The frequency spectrum may be truly broadband for this PEG category. Thus, random environmental vibrations will not have the same effect in both cases.

## 2.2. Piezoelectric material

Choice among many available piezoelectric materials is mainly guided by the need of an important intrinsic electromechanical coupling, leading to a high power density and facilitating PEG miniaturization. Material stiffness and maximum stress characteristics help in the design of the mechanical part. The material mechanical quality factor may also be also a parameter to consider, more particularly for power conversion in high frequencies. The output voltage range needed and the available stress levels help to determine the inter-electrodes distance. Most common geometries are plates or disks, with different possibilities for electrodes disposition. Electromechanical conversion is mainly governed by the longitudinal or transverse physical properties of piezoelectric material, respectively indexed 33 and 31, depending on the relative directions of the poling axis and the mechanical strain/stress axis.

Energy conversion analysis can be derived from piezoelectric equations (1) written according to IEEE standards, where $\{T\}$, $\{S\}$, $\{E\}$ and $\{D\}$ are respectively the stress vector, the strain vector, the electric field vector and the electrical induction vector. The piezoelectric material physical properties are the elastic stiffness matrix determined at constant electric field $[c^E]$, the piezoelectric stress matrix $[e]$ and the permittivity matrix at constant strain $[\varepsilon^S]$. '$t$' refers to matrix transpose. Depending on the piezoelectric insert geometry and coupling used, either axial or lateral strain/stress as well as the proper elastic and piezoelectric coefficients may be considered, leading to the simplified scalar expressions (2).

$$\begin{Bmatrix} T \\ D \end{Bmatrix} = \begin{bmatrix} c^E & -e \\ e^t & \varepsilon^S \end{bmatrix} \begin{Bmatrix} S \\ E \end{Bmatrix} \qquad (1)$$

$$\begin{cases} T = c^E S - e\, E \\ D = e\, S + \varepsilon^S E \end{cases} \qquad (2)$$

For frequencies lower than the proper resonances of the piezoelectric insert, equations (2) can be expressed as a function of the average displacement $u$, the piezoelectric voltage $V$, the equivalent mechanical force $F$ and the outgoing electric charge $Q$, leading to equation (3). Equations (4) define the clamped capacitance $C_0$, the force factor $\alpha$, and the short-circuit stiffness $K^E$ of the piezoelectric insert. $t_p$ is the distance between electrodes and $w_p$ is the insert width in the direction of the considered strain, $A$ is the surface of an electrode.

$$\begin{cases} F_P = K^E \cdot u + \alpha \cdot V \\ Q = \alpha \cdot u - C_0 \cdot V \end{cases} \qquad (3)$$

$$C_0 = \frac{\varepsilon^S A}{t_p}, \quad \alpha = e \cdot \frac{A}{t_p}, \quad K^E = c^E \cdot \frac{A}{w_p} \qquad (4)$$

Thus, from the electrical point of view, the second equation (3) shows that the piezoelectric insert used in a PEG can be basically modeled by a current source controlled by the mechanical velocity $I_0 = \alpha \cdot \dot{u}$, in parallel with a capacitor $C_0$.

### 2.3. Power optimization interfaces

As stated in the beginning of Section 2, the AC-DC power converter represented on Figure 1 can be used as power optimization. A first power optimization approach, so called "Standard" approach, consists to considering the structure of Figure 2, where the AC-DC power converter in composed of a diode rectifier and a smoothing capacitor $C_R$. The electric load is modeled by an equivalent resistor $R_L$. The corresponding displacement, voltage and current waveforms are represented on Figure 3. For this configuration, expression (5) gives of the average electrical power produced by the PEG in the case of an harmonic mechanical displacement $u$, whose amplitude and angular frequency are $U_M$ and $\omega$ [2,3,8].

$$P = R_L \frac{(\alpha \cdot \omega \cdot U_M)^2}{\left(R_L C_0 \omega + \frac{\pi}{2}\right)^2} \quad (5)$$

A thorough analysis would require a model of the mechanical part of the considered PEG. Indeed, the mechanical energy converted by the active material may induce more or less important variations on the mechanical displacement $u$ and thus on the generated current $I_0$. But considering a weak mechanical damping effect, the PEG power reaches a maximum $P_{Max}$ for a particular value $R_{Lopt}$ of the load equivalent resistance (6).

$$P_{\max} = \frac{\alpha^2 \omega}{2\pi C_0} U_M^2, \qquad R_{Lopt} = \frac{\pi}{2 C_0 \omega} \quad (6)$$

Considering the above equations, in the case of harmonic vibrations the PEG power optimisation can be summarized as a tuning between the displacement amplitude $U_M$ and the load resistance $R_L$. In an actual PEG, the load may include an energy storage cell such as a large capacitor or an electrochemical battery to overcome variations of ambient vibrations. In this case, a constant load voltage $V_R$ is considered instead of the load equivalent resistance $R_L$. The PEG power expression is then given by (7). The PEG maximum power $P_{Max}$ is the same as in (6), but it is expressed in (8) as a function of the optimal load voltage $V_{Ropt}$ related to the displacement amplitude.

Maximization of the PEG power whatever the mechanical displacement amplitude requires an additional interface shown on Figure 4, allowing a permanent tuning of the rectified voltage $V_R$ to the optimal value. This interface, a DC-to-DC power converter, must induce as few power losses as possible.

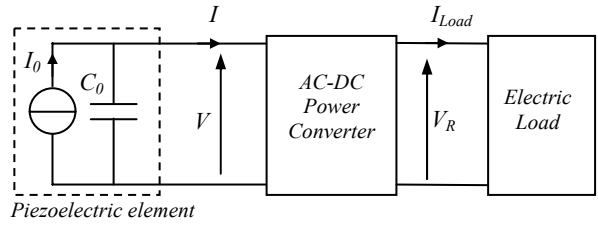

Figure 1: PEG including an AC-DC power conversion interface

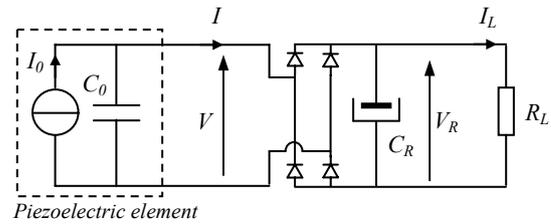

Figure 2: PEG using a diode bridge as an AC-DC power conversion interface

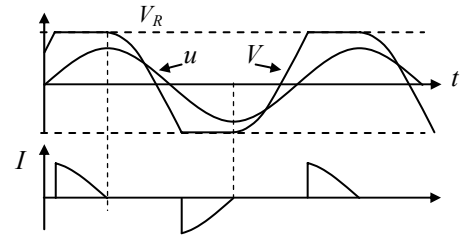

Figure 3: Displacement, piezoelectric voltage and current

$$P = \frac{2C_0 \omega}{\pi} V_R \left(\frac{\alpha}{C_0} U_M - V_R\right) \quad (7)$$

$$P_{Max} = \frac{2C_0 \omega}{\pi} \cdot V_{Ropt}^2 \qquad V_{Ropt} = \frac{\alpha}{2C_0} U_M \quad (8)$$

In this domain, switching-mode power converters are known for their high efficiency. The significant development of portable electronic devices in the past years led to important technological progress of low-power switching-mode power converters: their efficiency reaches now currently the region of 90%-95%. This first power optimization approach may be qualified of "quasi-linear", although the diode rectifier bridge and the smoothing capacitor induce a distortion on the piezoelectric voltage $V$. But previous works showed that some DC-DC power converters may also be used to increase in a particular way the piezoelectric voltage nonlinearities [7,8]. As a result, the energy conversion capability of the active material is greatly enhanced. Another advantage of this nonlinear technique, so called "Synchronous Electric Charge Extraction" (SECE), is its self-optimization property: the PEG power remains at maximum whatever the load voltage or equivalent resistance.

The SECE technique is implemented with the buck-boost power converter structure represented on Figure 5, with a particular control: the transistor $T$ is turned ON each time the rectified voltage $V_R$ reaches a maximum, and it is turned OFF each time $V_R$ reaches zero. So, at each triggering time, the power converter removes completely the electric charge stored on the piezoelectric element electrodes and transmits the corresponding energy to the load, through the inductor $L$. The inductor value is chosen in considering the piezoelement capacitance, so that the time needed to extract the electric charge is much shorter than the vibration period. The corresponding mechanical displacement, piezoelectric voltage and current waveforms are presented on Figure 6. The average power converted by the active material is given in (9). According to equations (6) and (9), with the same mechanical displacement amplitude the power converted is four times greater using this technique than with the firstly presented power optimization approach.

$$P = \frac{2\alpha^2 \omega}{\pi C_0} U_M^2 \qquad (9)$$

### 3. MULTIMODAL VIBRATIONS

Starting from the power optimization analysis in the case of harmonic mechanical vibrations presented in Section 2.3, it is possible to consider the PEG behavior in the case of steady state, multimodal mechanical vibrations. The mechanical part of the considered multimodal PEG is a cantilever steel beam, whose three first bending modes frequencies are respectively 56 Hz, 334 Hz and 915 Hz. The beam vibrations are driven by an electromagnet controlled by a function generator through a power amplifier. The active material is P1-89 PZT ceramic (Saint-Gobain). This piezoelectric insert, composed of six 9x28x0.3 mm$^3$ plates, is bonded near the clamped edge. The capacitance value of $C_0$ is 41.8 nF. A second identical piezoelectric insert, bonded on the opposite side of the cantilever beam, is used as mechanical displacement transducer. Both power optimization circuits are used: the "Standard" one is composed of a diode bridge, and a 2.2 µF smoothing capacitor in parallel with a variable resistor; the SECE is implemented as the circuit represented on Figure 5, but the electrochemical battery is replaced by a 2.2 µF capacitor. The control circuitry of this second optimization interface is powered by an external power supply. First, an experimental analysis of the PEG behavior is made for each of the mechanical modes separately. For each frequency, the displacement amplitude is tuned in order to reach a 1 mW maximum power for the Standard interface.

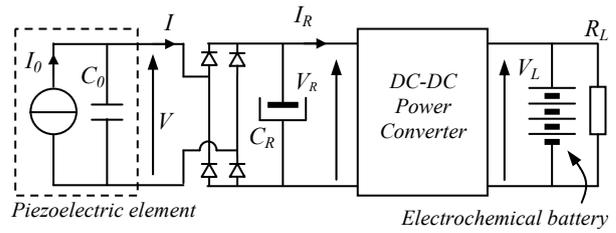

Figure 4: PEG including a DC-DC power converter for power optimization

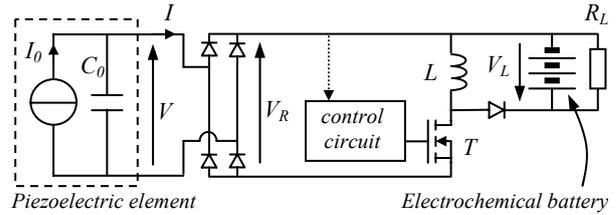

Figure 5: Synchronous electric charge extraction circuit

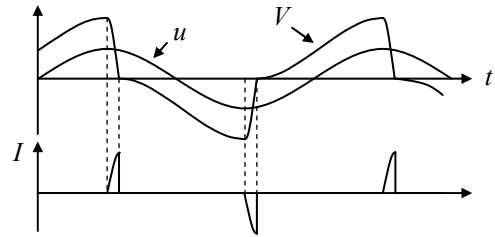

Figure 6: Displacement, piezoelectric voltage and current in the case of synchronous electric charge extraction

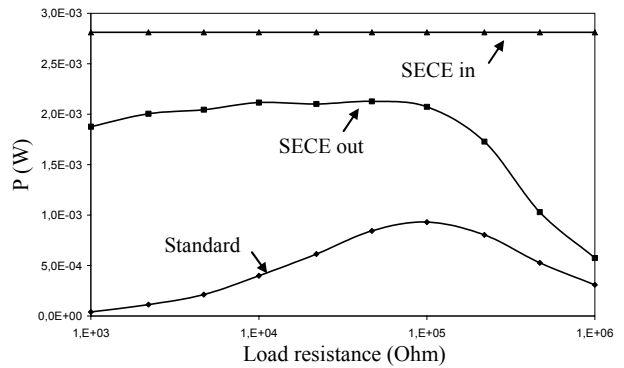

Figure 7: First mode only (56 Hz)

Experimental results presented on Figure 7, Figure 8 and Figure 9 confirm that maximum power of the Standard technique is reached with the optimal load resistance values defined in equation (6), which are respectively 106 kΩ, 18 kΩ and 6.5 kΩ. Plots referred as "standard" and "SECE in" represent the average power outing from the rectifier bridge respectively for the Standard interface and the SECE interface.

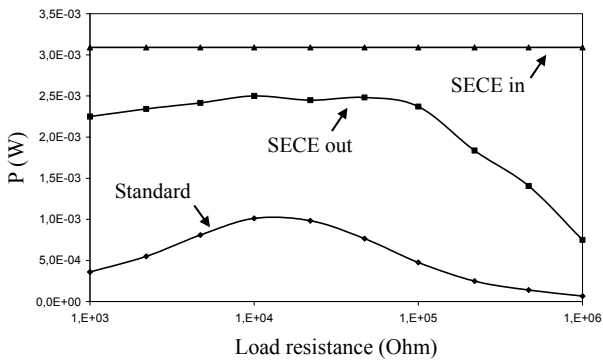

Figure 8: Second mode only (334 Hz)

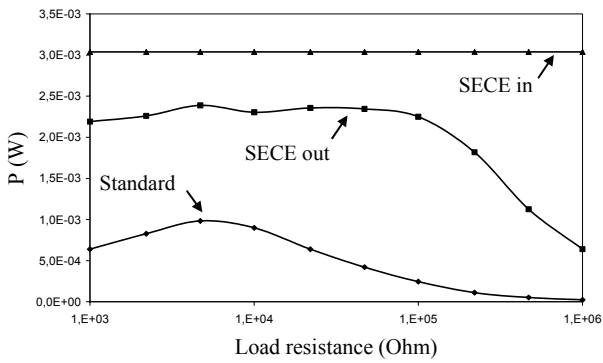

Figure 9: Third mode only (915 Hz)

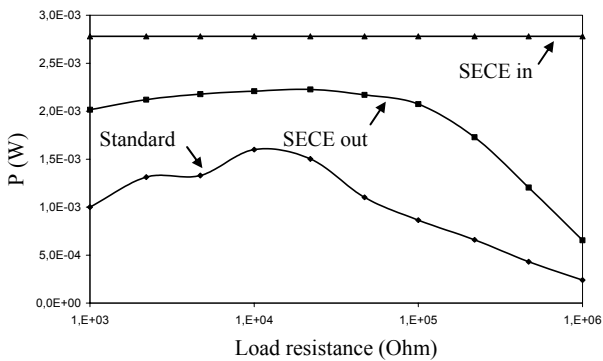

Figure 10: Three modes mixed

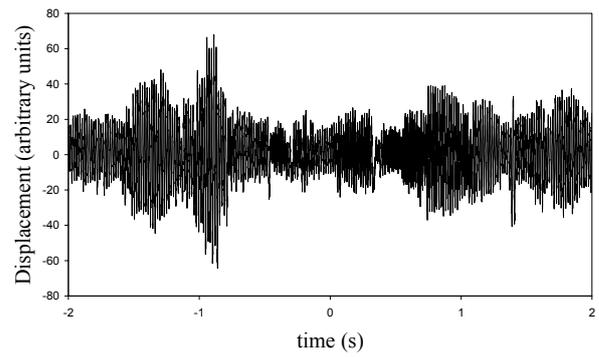

Figure 11: Random displacement as a function of time

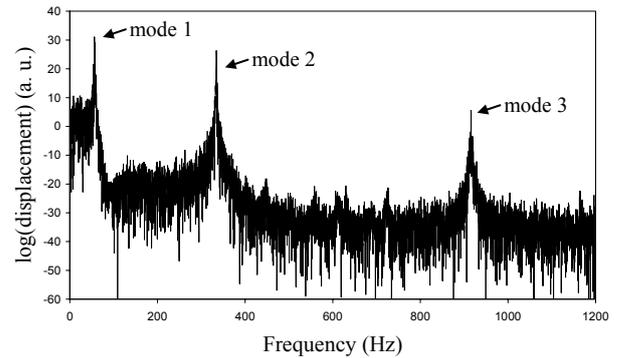

Figure 12: Random displacement spectrum

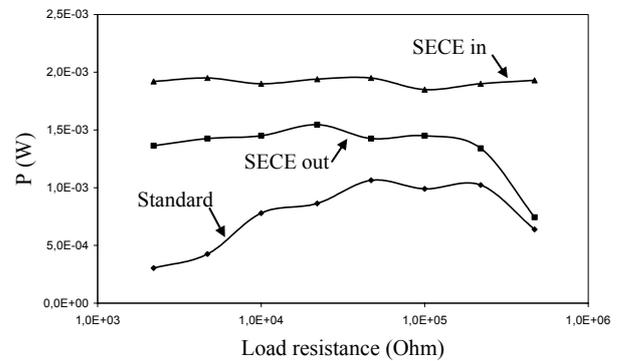

Figure 13: Powers as a function of the load resistance, PEG powered by random vibrations

Plots referred as "SECE out" represent the average power of the terminal load resistance with the SECE interface. According to (6) and (8), the SECE technique should theoretically harvest four times more power than the Standard technique, but measurements show that the power is only increased by a factor three. Difference between theoretical and experimental power gains may be in particular due to the piezoelectric material leakage resistance which is not taken into account in the model.

Input power of the SECE circuit ("SECE in") confirms to be constant whatever the load resistance, whereas output power ("SECE out") decreases for high values of the load resistance because of the circuit imperfections. The efficiency of the circuit is practically 80% in the range 1 kΩ to 100 kΩ, the consumption of the control circuitry not being taken into account.

When several modes are mixed, experimental results show that the power harvested with the Standard circuit remains smaller than with the SECE circuit, but the power ratio is weaker than in the case of harmonic vibrations. Figure 10 shows for instance a case in which the maximum power with the Standard interface is 1.63 mW whereas the input power of the SECE interface is 2.76 mW: the power gain is here 1.7 only. In some cases, the SECE technique may lead to harvested powers lower than the standard technique, typically in the case of PEG vibrations composed of a low-frequency mode with large amplitude, mixed with a high-frequency mode with very weak amplitude. This drawback may be solved by selecting the right triggering times for extracting the electrical charge stored on the piezoelectric element, that is to say making the good choice among the local extrema of the piezoelectric voltage.

## 4. RANDOM VIBRATIONS

The random force driving the PEG cantilever beam is generated by a HP 35665A random noise generator, powering the electromagnet through a power amplifier. An image of the mechanical displacement measured with a piezoelectric insert bonded on the PEG cantilever beam near the clamped end is shown on Figure 11. The displacement spectrum plotted on Figure 12 reveals the mechanical filtering of the cantilever beam: the mechanical vibration is mainly composed of the three bending modes, respectively at 56 Hz, 334 Hz and 915 Hz. The average harvested powers presented on Figure 13 are the average values of measurement performed during 100 seconds. The power harvested with the SECE technique ("SECE in") is here roughly twice the maximum power harvested with the Standard technique.

## 5. CONCLUSION

The results presented in this paper show that it is possible to harvest effectively broadband vibrational energy, having random variations. Comparison between the Standard power optimization technique and the so called "Synchronized Electric Charge Extraction" technique exhibit several advantages of this last one. First, there is no need for load impedance adaptation. This property represents a great advantage, because tracking the optimal load resistance value needed to optimize the Standard interface power may be relatively complicated in the case of broadband, random vibrations. Then, for a given mechanical vibration the power harvested by the new interface may be increased by a factor up to 400%. Moreover, practical implementation of the power optimization technique is much simpler than in the case of the Standard technique.

Ongoing works aim at improving the efficiency of electronic circuits for both techniques, in order to make comparisons of completely self-powered devices.

Another part of these research works will be focused on the development of smart strategies for selecting the right mechanical displacement extrema as triggering instants for electric charge extraction, in order to improve the broadband performance of the new technique.